# Convolutional Neural Networks based automated segmentation and labelling of the lumbar spine X-ray

Sándor Kónya [1*], Sai Natarajan T R [2], Hassan Allouch[3], Kais Abu Nahleh[3], Omneya Yakout Dogheim[1], Heinrich Boehm[3]

1: Zentralklinik Bad Berka, Center for Diagnostic & Interventional Radiology and Neuroradiology, Bad Berka, Germany
2: Independent Researcher, Chennai, India
3: Zentralklinik Bad Berka, Department of Spinal Surgery, Bad Berka, Germany

**Correspondence:**
Dr. med. Sándor Kónya
Sandor.Konya@zentralklinik.de

**Abstract**

The aim of this study is to investigate the segmentation accuracies of different segmentation networks trained on 730 manually annotated lateral lumbar spine X-rays. Instance segmentation networks were compared to semantic segmentation networks. The study cohort comprised diseased spines and postoperative images with metallic implants. The average mean accuracy and mean intersection over union (IoU) was up to 3 percent better for the best performing instance segmentation model, the average pixel accuracy and weighted IoU were slightly better for the best performing semantic segmentation model. Moreover, the inferences of the instance segmentation models are easier to implement for further processing pipelines in clinical decision support.

**Disclosures**:

The present institutional, retrospective analysis was conducted in accordance with institutional and national ethical guidelines and thus in accordance with comparable ethical standards to the 1964 Helsinki Declaration[1]. The study was not supported by any research grant.

**Conflict of Interest**

The authors declare that they have no conflicts of interest.

## Introduction

Semantic segmentation algorithms are a subset of supervised learning algorithms that control the training of Convolutional Neural Network classifiers from a set of labelled training image data. These classes of algorithms produce class labels for every pixel in an input image. From a bird's eye view, this task can be mathematically formulated as: given a 2-dimensional input array of NxM pixels of an image p= [$p_{11}$, $p_{12}$, ..., $p_{1M}$; $p_{21}$, $p_{22}$, ..., $p_{2M}$;...;$p_{N1}$, $p_{N2}$, ...,$p_{NM}$], the classifier's job is to predict the per pixel class label from a set of labels L= [$l_0$, $l_1$, ..., lp].

In the recent past, a number of semantic segmentation models have come out, competing with each other, trying to overcome the shortcomings of their counterparts. Even though these approaches are very promising, it becomes difficult at one stage to find out which one among the many is suited best for a specific research area. Also, most of the semantic segmentation models have been tested for the popular datasets such as cityscapes, coco, pascal, etc. For the purpose of our study, we have selected a few state-of-the-art semantic and instance segmentation algorithms and compared their performance using various factors and criteria in the domain of medical image analysis.

In the era of MRI and CT-scans, X-rays are still one of the most commonly requested diagnostic imaging modalities, accounting for 1,7 procedures on each inhabitant in Germany per year[2]. The recognizable pathologies of the lumbar spine on a mere X-ray have a wide spectrum from developmental anomalies through acute traumatic conditions to degenerative disorders. In order to create an automated system that performs morphometric analysis of the X-rays, we need a proper segmentation of the objects on the radiographs, like vertebrae and in case of postoperative images various implants like screws or cages.

The analysis of lumbar X-rays involves not only the classification on the present pathologies but also the measurement of various angles (such as angle of lordosis) and geometric distances. The measurements rely on the localization and the position of each vertebra compared to each other. There are degenerative conditions that modify the overall appearance of the vertebral body, making it difficult to delineate the contour of the vertebra such as fractures but also osteophytes, that are present in almost every moderately or severely degenerated segment. This is why we need an automated system that performs good segmentation of the vertebra instances and other objects even and especially in images reflecting severe deformities. We compared the performance of recent neural networks to find the best suitable model for this purpose.

## Material

For the training, a set of 730 lateral lumbar spine radiographs were selected and retrieved from the PACS (Picture Archiving and Communication System) with the pynetdicom[3] library running in a portable local WinPython[4] environment inside a workstation and 100 further images were annotated for testing purposes. The dicom images contained preoperative and postoperative studies, before and after mono- or multi-segmental interbody fusion and posterior instrumentation. Indications for surgery were underlying severe segmental degenerative disc disease with or without segmental instability. Therefore, on the images, at least one but usually multiple severely diseased intervertebral spaces were present. The annotations were performed using the open-source VIA (VGG Annotation tool[5]) by trained radiologists and spine surgeons having experience of more than 5 years. Each vertebra was labelled with a multi-point polygon. Cages, screws, and instrumentation were also labelled. The authors chose to create 2 distinct classes, one for lumbar and one for sacral vertebrae instead of 6 separate classes for each lumbar vertebra, due to the similarity between the lumbar vertebrae and the differently shaped sacral vertebrae S1 and S2. The resulting annotations were exported in the JSON format and further processed in Jupyter notebook with pandas and OpenCV.

**Methods**

The authors used the anaconda environment, which is an open-source distribution environment of the Python programming language. Different libraries and their dependencies were installed such as Keras, TensorFlow or PyTorch. In this environment, a pre-installed Jupyter Notebook was used as the development platform. Different semantic- and instance segmentation algorithms were compared for the domain-specific task of automated segmentation and labelling of lumbar spine X-ray. The models had not only different internal loss function and internal metrics, they had different prediction formats, so the authors decided to compare the models with common metrics based on the merged binary masks of the model predictions.

**Workflow**

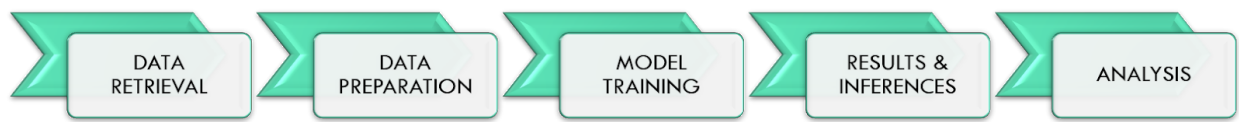

(Figure 1.)

**Data Retrieval:**

As shown in Figure 1., data for the project was obtained directly from the Picture Archiving and Communication System running on the server. Around 730 images were retrieved, anonymized and annotated manually from the archive and another 100 for testing purposes.

**Data Preparation:**

Each model required slightly or very different input formats (JSON, image masks, raw image material) and the resolution constraints also had to be fixed, which are features and special requirements of each model. For further information see references for each model.

**Model Training:**

Once the datasets were prepared, the next stage in the pipeline was to start the training of the segmentation models. The authors used a GeForce RTX 2080 Ti graphic card for the training. The dataset was split into 90% for the training set and the remaining 10% for the validation. We have selected 5 different segmentation models for the training of the dataset. They are categorized into Semantic and Instance Segmentation models. UNet, PSPNet, and Deeplab v3 are semantic segmentation models whereas MaskRCNN and YOLACT are instance segmentation models. A brief introduction about all the models is given below:

**U-NET[6]**

We included the game-changing biomedical image segmentation model, published in 2015 since it has a proven record of performing extremely well when given limited training data. Our analysis was based on the Keras implementation[7] of the code. It is a standard Convolutional Neural Network model that is widely used for segmentation related tasks especially in medical imaging such as nerve classification, cell segmentation, etc..

**Mask R-CNN[8,9]**
Mask R-CNN is an extended Faster R-CNN[10] with an additional branch for predicting segmentation masks on each Region of Interest (RoI), in parallel with the existing branch for classification and bounding box regression. The side branch predicts a segmentation map for each RoI. MaskRCNN has a good history with its predecessors. It uses selective search (used in RCNN) to choose region proposals in an image and then uses a CNN to extract a 2048 feature vector. This vector is passed on to a linear-SVM classifier. The MaskRCNN also extends the Faster-RCNN by adding an additional module that performs segmentation.

**PSPNet[11,12]**
Pyramid Scene Parsing Network, was the champion of the ImageNet Scene Parsing Challenge in 2016. It uses a Pyramid Pooling Module, with aggregated context from different image regions by pooling the feature maps from coarse (global average pooling) to the fine (6x6) regions. Feature maps in PSPNet are extracted using the Dilated Network (Atrous Convolution) of a pre-trained Residual Network (ResNet) model. The feature maps are further fed to the Pyramid Pooling Module in order to different patterns. The output from the Pyramid Pooling Module is up-sampled and stacked together to get the feature maps. This feature map is then passed to the Fully Connected Layer to generate segmentations.

**YOLACT (You Only Look At Coefficients)[13]**
This is a simple, fully-convolutional model that was designed for fast instance segmentation. The real-time performance is achieved by breaking the instance segmentation into two parallel tasks: a Prototype Generation Branch and a Mask Coefficient Branch that is combined using a linear combination of the former with the latter as coefficients to assemble the final mask.

**DeepLabV3[14]**
This model uses an atrous spatial pyramid pooling (ASPP) to robustly segment objects at multiple scales with filters at multiple sampling rates and effective fields-of-views, augmented with image-level features to capture context information.

**Results and Inference:**

Once all the models had completed training (training time varied from 10 - 48h), the predictions were made for the test set. We had tested each model one after another and obtained different metrics for comparison studies which are described further down in this article.

**Metrics used:**

In order to evaluate the segmentation models, we have used different performance metrics inspired by Long et al.[15] from the repository of Martin Keršner[16]. Some of the metrics are purely domain-specific and we have also taken other common metrics that are used for the evaluation of segmentation models.

  a. **Pixel Accuracy:** Pixel accuracy can be calculated as the percentage of pixels in

the generated labelled mask that are classified correctly when compared with the original labelled mask.
b. **Intersection over Union (IoU):** The most common metric for the evaluation for segmentation models is the IoU. IoU is also known as the Jaccard Index. IoU can be calculated as the overlapping area between the predicted labelled mask and the original labelled mask (Figure 2.). For a multi-class segmentation such as ours, the mean-IoU is calculated by averaging the IoU of each class.

(Figure 2.)

c. **Mean Accuracy:** Mean Accuracy is defined as the mean of no. of pixels of class $i$ predicted to class $i$ over the total no. of pixels of class $i$ for all classes $i$.
d. **Frequency Weighted IoU:** In frequency weighted IoU, the IoU for each class is computed first and then average over all the classes is computed.

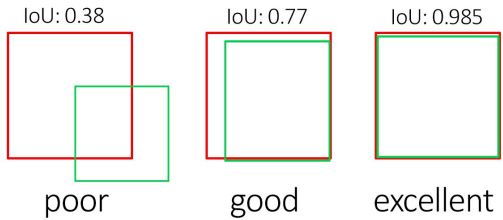

poor good excellent

**Results**

|  | U-Net | Mask R-CNN | PSPNet | DeepLabV3 | YOLACT |
|---|---|---|---|---|---|
| **Pixel Accuracy Average** | **98.17** | 96.58 | 97.88 | 98.00 | 97.84 |
| **Mean IoU Average** | 88.64 | 86.78 | 86.50 | 88.14 | **91.64** |
| **Mean Accuracy Average** | 92.65 | 90.25 | 90.91 | 92.25 | **94.43** |
| **Frequency Weighted IoU Average** | **96.48** | 93.47 | 95.93 | 96.16 | 95.84 |

(Table 1.)

The results clearly show that the best instance segmentation (YOLACT) and best semantic segmentation (U-Net) model delivers very close metrics values in terms of average pixel accuracy and frequency weighted average IoU. More importantly a relatively large, up to 3% difference in average mean IoU and 1.8% by average mean accuracy (Table 1.) in favour of YOLACT can be observed. Figure 3 shows how huge alterations in the visual appearance can hide behind the seemingly low mathematically differences. In Figure 3. we show two cases: **(A)** where the semantic models deliver better than average pixel accuracies, and Mask R-CNN due to a falsely labelled extra "vertebrae" lowers its scores as an example for the sources of failure. In case **(B)** all models but YOLACT deliver an under average pixel accuracy, whereas the reasons are two-fold: first, there are two overlapping vertebrae that had been recognized correctly in the instance segmentation models, and as one large fused blob in the semantic models. Secondly an extra, thoracic vertebra (11th thoracic vertebra) that had been at least partially delineated on all models but YOLACT. Only vertebrae as Th 12 or caudally to it are present on the ground truth masks. In consequence, if there are more vertebrae in the predicted mask than in the ground truth mask, it lowers the accuracy.

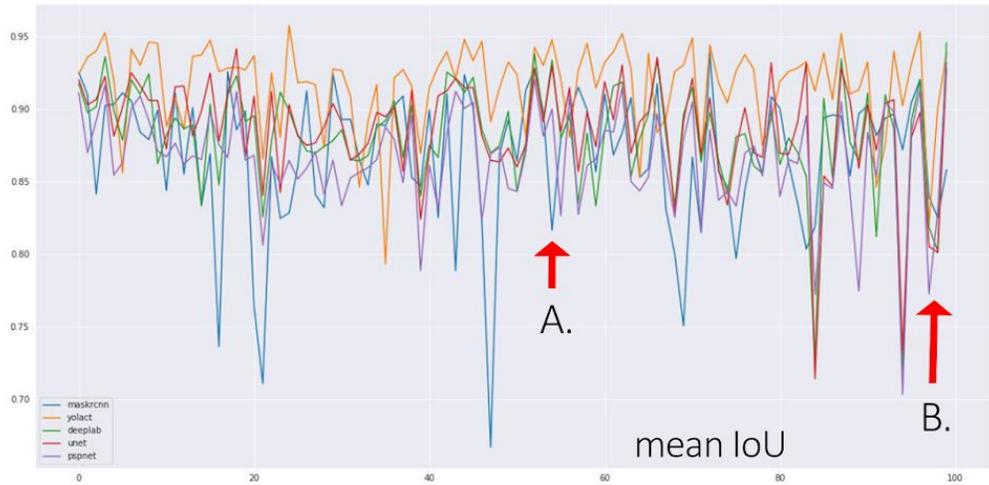
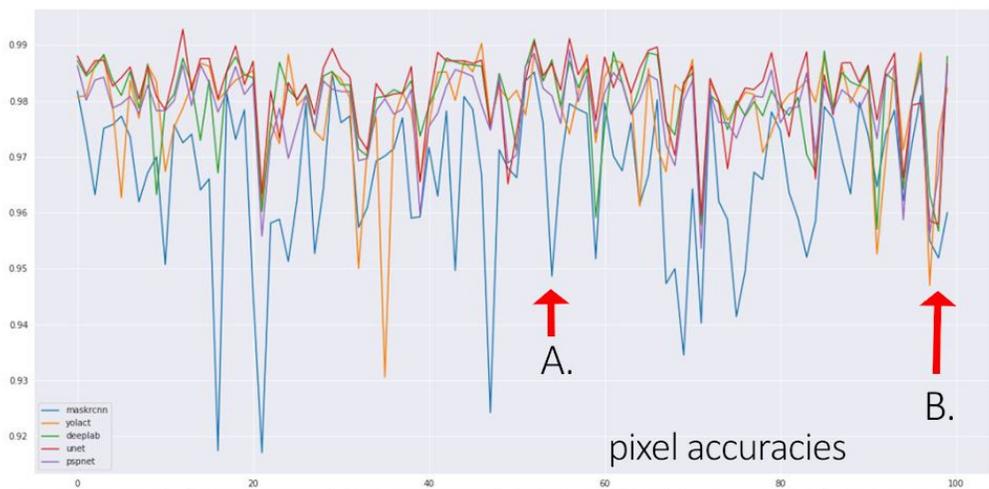
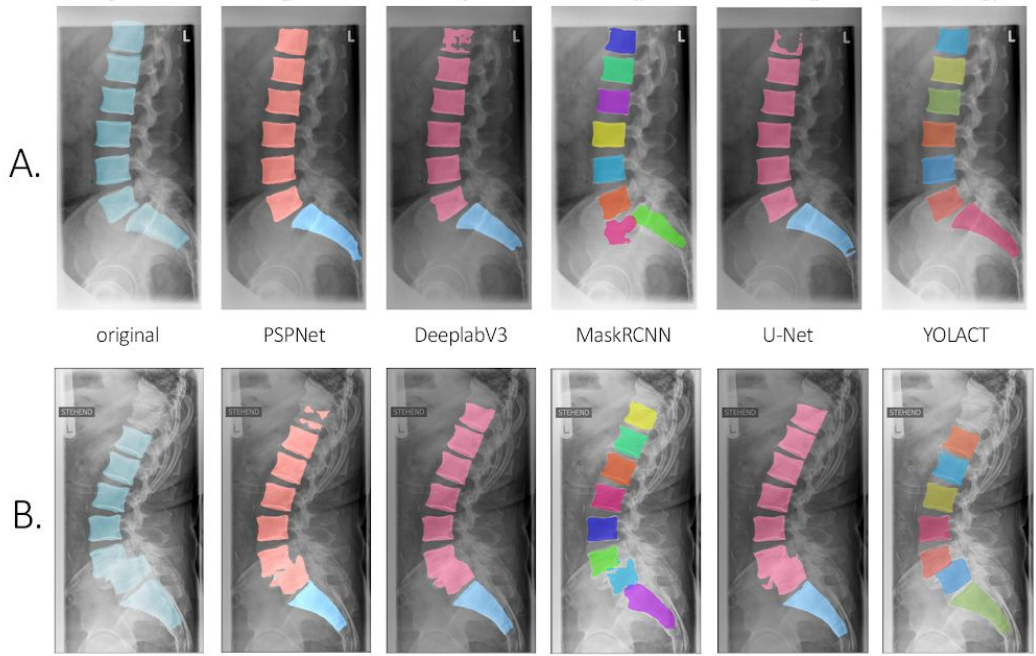

(Figure 3.)

## Discussion

The recent development of open-source segmentation networks has enabled researchers to experiment with the automation and analysis of biomedical images[17]. Thus creating semi- or fully automated image processing workflows for segmentation is fast and feasible. Since the internal metrics of each model are based on slightly different methods, for the current study a common metric was chosen based on a merged binary mask of the prediction. This method results in information loss in case of instance segmentations where the predicted separate masks of each instance may overlap in the lateral projection, appearing as fused vertebrae on the binary mask and also by the metallic implants, cages that usually partially overlap with the vertebral borders. Even with this simplification the YOLACT based instance segmentation model outperforms the semantic segmentation models regarding average mean IoU while pixel accuracies are close to each other. A class-based segmentation accuracy was not within the scope of this article but poses further research direction. The subjective impression that the contours of the vertebrae are wavy by the YOLACT model (A.) and smooth by the U-Net model (B.) makes the segmentation mask though mathematically correct, visually somewhat distractive (Figure 4.).

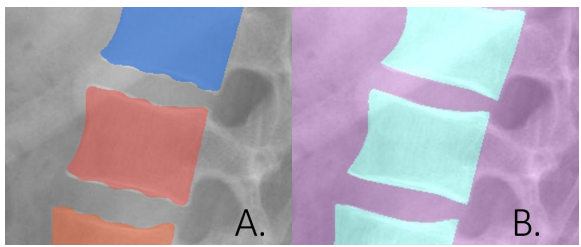

(Figure 4.)

The human-level spinal vertebra segmentation is the first and essential step for many computational spine analysis tasks as for example different spatial relations of vertebral bodies (Cobb angle or lordosis angle) and shape characterization. There are only a few published articles that utilize the power of Convolutional Neural Networks in the segmentation of spine X-rays; Recently Cho et al.[18] utilized a U-Net based model with very good performance on segmenting the lumbar vertebrae on their own set although there were some very important limitations. First of all they did not consider to include any kind of implants such as cages, which limits the utility of the system for practical day to day purposes in outcome evaluation of surgical treatment. Further limitation is that no severe deformities of the vertebral bodies and intervertebral spaces were included. This implies that their algorithm is not applicable for the majority of the patients. In our study we overcome these limitations by training on a sample containing over 70% of images depicting implants (multiple cage systems, screws, rods and also cement augmentations). In addition to that we included the full spectrum of degeneratively diseased intervertebral spaces. An interesting feature found during testing different images was that the YOLACT model was able to partially identify beforehand unseen and as such untrained objects, such as vertebral body replacement, however with overlapping labelling and false annotation (Figure 5.).

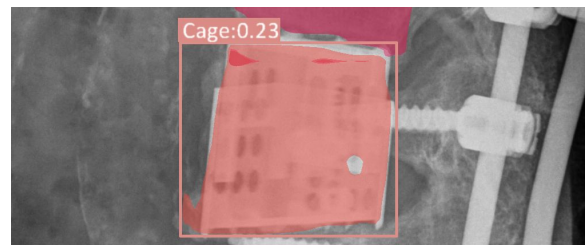

(Figure 5.)

Another surprising finding was the ability of the YOLACT model to delineate fractured vertebrae, even though these were also not part of the training set (Figure 6).

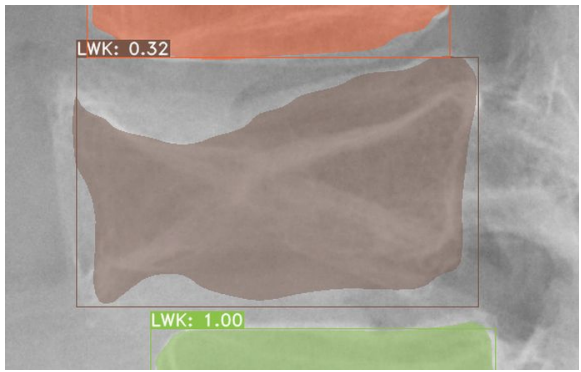

(Figure 6.)

Kuok et al.[19] used a three-step system that included spine RoI detection, vertebrae RoI detection and vertebrae segmentation on a.p. whole-body images. The image processing steps helped to find the RoIs for each vertebra and a U-Net performed the segmentation of the vertebrae instances within the RoI. The image segmentation was thus performed on a cropped image region that helps to get better pixel accuracy. In our study, the instance segmentation models accepted the full resolution images and we performed no RoI cropping. Sa et al.[20] used Faster-RCNN object detection as the first step towards automatically identifying landmarks from spine X-ray images. In their work intervertebral spaces were detected without the contour detection of the vertebrae themselves. Although our study focuses on the detection of the contour of the vertebrae, the endplates draw the exact borders of the intervertebral spaces. So the extraction and analysis of these can be simply derived from the predicted masks.

The fact that the best-ranking instance segmentation model performed better than all other semantic segmentation models is fortunate since the results of an instance segmentation model can be used without any extensive post-processing step, given no vertebrae is missed and the class accuracy is good. Removing overlapping class predictions with non-maximal suppression is possible.

The semantic segmentation masks contain fused binary masks of multiple vertebra instances. The pixels of one class are not separated onto instances, that makes the labelling of each vertebra instance difficult and in every case needs morphological postprocessing, especially in severe degenerative cases or when cages are present (Figure 7.).

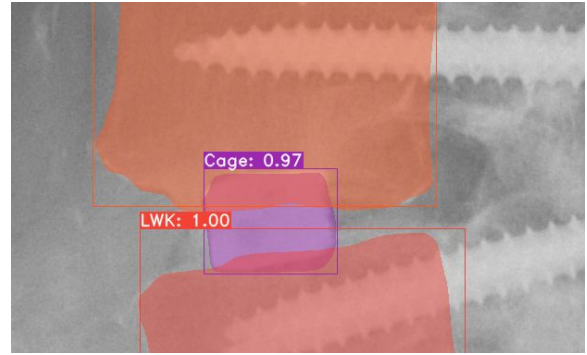

(Figure 7.)

The postprocessing of a semantic segmentation model is nontrivial and was not addressed in this article. Special cases may arise in instance segmentation when the segmentation model skips predicting an instance (due to low prediction score) or predicts multiple overlapping classes or instances. The first case can be addressed by lowering the threshold score for the prediction. In case of multiple overlapping instances the prediction with the higher score can be chosen with non-maximum suppression[21]. This way the labelling of each vertebra could be easily performed based on the order and distance from the sacral vertebra. With further conventional morphological postprocessing we are able to analyse postoperative images. Every degeneration or pathological alteration that causes morphological change of the anatomical structures can be further processed. This is promising in regard to changes in intervertebral space, osteophyte formation, dislocation or failure of metallic implants, especially since it allows analysis of those features over time. On Figure 8. possible steps of information extraction are shown, A. contour segmentation, B. osteophyte and corner detection with erosion,

dilation and subtraction, C. endplate approximation.

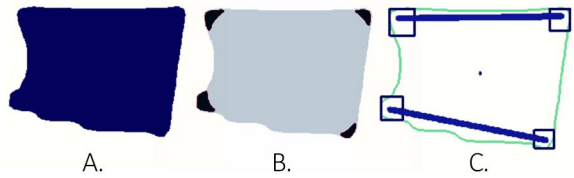

(Figure 8.)

We observed that the more severe degenerations are present, the better is the prediction of the instance segmentation models (probably due to the correct identification of the overlaps) and in the case where less degenerative changes are present, the semantic segmentation models achieve better accuracies. This shows an interesting direction for further research but is well beyond the scope of the present study.

**Conclusion**

In this article, we have compared state of the art semantic segmentation and instance segmentation models for automated segmentation of vertebrae in standard lateral X-rays of a real-life clinical cohort with severe degenerative diseases and metallic implants. The best instance segmentation model yields better average mean IoU values and better mean average accuracy than the best semantic model. It also allows better segmentations of overlapping vertebrae. The results can be used for creating further clinical decision support pipelines without much morphological postprocessing.

**Author contributions**

SK and SN: carried out the experiments of the project and did the coding. HB, AH, and KAN: provided guidance on the methodology and overall project. OYD, KAN, and SK: provided technical support in terms of the annotation process. HB, AH, SK: generated research ideas and reviewed the manuscript.